# Seismic Site Effects for Shallow and Deep Alluvial Basins: In-Depth Motion and Focusing Effect

J.F. Semblat, P. Dangla, M. Kham, Laboratoire Central des Ponts et Chaussées,
58, bd Lefebvre, 75732 PARIS Cedex 15, France, semblat@lcpc.fr
A.M. Duval, CETE Méditerranée, Seismic Risk Team, Nice, France

**Abstract** : The main purpose of the paper is the analysis of seismic site effects in various alluvial basins. The analysis is performed considering a numerical approach (Boundary Element Method). Two main cases are considered : a shallow deposit in the centre of Nice (France) [1] and a deep irregular basin in Caracas (Venezuela) [2].

The amplification of seismic motion is analysed in terms of level, occuring frequency and location. For both sites, the amplification factor is found to reach maximum values of 20 (weak motion). Site effects nevertheless have very different features concerning the frequency dependence and the location of maximum amplification. For the shallow deposit in Nice, the amplification factor is very small for low frequencies and fastly increases above 1.0 Hz. The irregular Caracas basin gives a much different frequency dependence with many different peaks at various frequencies. The model for Caracas deep alluvial basin also includes a part of the local topography such as the nearest mountain. One can estimate seismic site effects due to both velocity contrast (between the basin and the bedrock) and local topography of the site.

Furthermore, the maximum amplification is located on the surface for Nice, whereas some strong amplification areas also appear inside the basin itself in the case of Caracas. One investigates the influence of this focusing effect on the motion vs depth dependence. This is of great interest for the analysis of seismic response of underground structures. The form and the depth of alluvial deposits are then found to have a great influence on the location of maximum amplification on the surface but also inside the deposit for deep irregular basins. It is essential for the analysis of the seismic response of both surface and underground structures.

## 1. Introduction

The analysis of seismic site effects considers amplification versus frequency curves showing the range of the spectrum leading to large motion amplification. Experimental measurements are generally performed along the surface with various methods : microtremor recordings, real earthquakes measurements [3]. Information on in-depth motion could sometimes be obtained thanks to specific measurement networks [4,5]. Through numerical methods, one can also study the amplification process in various types of geological structures. It is for instance possible to consider the vibratory resonance of alluvial basin [6,7]. Otherwise, one can perform numerical analyses on site effects through explicit wave propagation models.

In this paper, we try to study the influence of the basin geometry on site effects. Both surface and in-depth motion are especially considered to find out how they can be modified by some specific motion amplification for a typical basin geometry. The focusing effects are for instance taken into account to explain the possible increase of in-depth motion in some areas [8]. To perform such an analysis, seismic wave amplification is investigated in various types of alluvial basins considering the boundary element method.

## 2. Shallow and deep alluvial basins

For the analysis of in-depth motion amplification and focusing effect, we chose two alluvial basins with very different profiles : the first one is located in the centre of Nice (France) and is a wide flat basin (width 2 km, depth 60 m) [1], the second one, located in Caracas (Venezuela), is a deep irregular valley surrounded by mountains (width 3.6 km, depth 300 m) [2]. Some experimental or numerical investigations were performed previously for both basins [1,2,3]. We found that the amplitude versus frequency dependence is very different in each case. We will then try to analyse the variations of in-depth motion in both cases and to find out if focusing effects can actually influence motion amplification in a deep irregular basin.

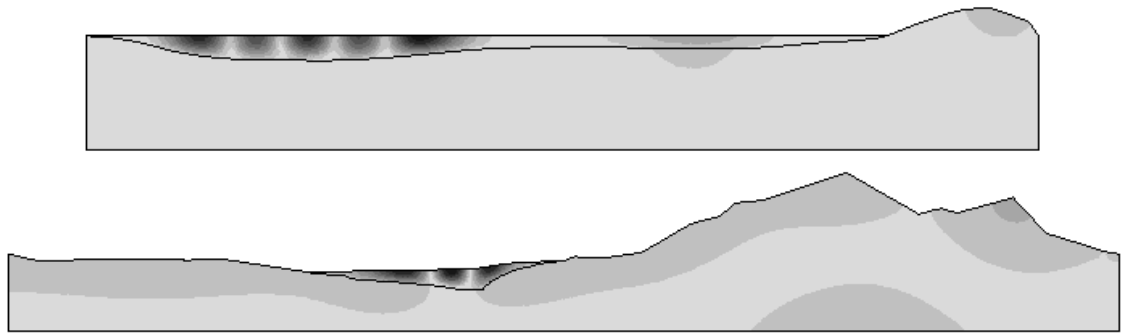

Fig.1 : BEM modelling of site effects for shallow and deep alluvial deposits :
amplification factor in the case of Nice (top) and Caracas (bottom)

## 3. Modelling site effects by the BEM

The numerical analysis of site effects for both types of basin was performed by the Boundary Element Method [1,2,9,10]. The method is very powerful since it allows the modelling of seismic wave propagation for large geological structures without such drawbacks as numerical dispersion for some other methods [11]. The numerical analysis was performed considering plane seismic waves of various types [10,13]. The shear wave velocities were chosen as follows: for Nice [1] $C_1$=300m/s in the deposit and $C_2$=1400m/s in the bedrock ; for Caracas [2] $C_1$=450m/s and $C_2$=2500m/s respectively. Fig.1 gives the isovalues of the amplification factor for both sites. The first one (Nice, French Riviera) is shallow and its geometry is very regular. Site effects are found to be strong in the deepest part of the deposit (left) between 1 and 2 Hz and in the thinnest part (right) for frequencies above 2 Hz [1]. For the second one (Caracas, Venezuela), there is a significant influence of the local topography (nearest mountains) as shown in Fig.1 for 0.6 Hz [2]. The surface motion amplification has a very complex dependence on frequency. The irregular form of this basin as well as the large velocity contrast suggest that focusing effects could occur in the basin itself and influence the amplification process. In the following, we will estimate in-depth motion variations to determine if they can lead to deep amplification areas. This issue is very important for the design of earthquake resistant underground structures [12].

## 4. Occurrence of focusing effect

Focusing effect is related to particular geological structures that can focus seismic energy because of their geometrical and mechanical features. Some unexpected localised zones of damage were especially observed after Northridge earthquake [8]. To analyse potential focusing effect, we compare the seismic motion amplification at various depths for both alluvial basins (shallow regular ; deep irregular). Fig.2 displays the amplification factor in the whole shallow basin (Nice) at various frequencies. It is given versus depth and distance (along the free surface). For the lowest frequency (1.0 Hz), there is only one amplification area on the free surface and in-depth motion decreases regularly. For frequencies values of 1.4 and 1.6 Hz, there are several amplification areas along the free surface in the left deepest part of the basin. No significant amplification is observed in the thinnest part (right) for those frequency values. For larger frequency values (2.0, 2.2 and 2.4 Hz), many different amplification areas are obtained along the free surface in the left part of the basin except for the last frequency value leading to low amplification in this part. In the right part, there is a strong increase of the amplification factor values for the three largest frequencies. Nevertheless, the seismic motion amplification is always decreasing with depth inside the basin. For all frequencies, there is a monotonic decrease of ground motion values from maximum surface motion values. Since the shallow basin is very flat, no focusing effects is observed but there is still a basin effect leading to amplification values much larger than those obtained from 1D analytical estimation considering the mechanical features of the deposit [1]. For the shallow regular basin in Nice, site effects are then influenced by basin effects leading to seismic waves trapped in the deposit. However, there is no energy focusing effect due to the basin geometry and consequently no large in-depth amplification.

In the case of the deep irregular basin in Caracas [2], amplification values versus depth and distance are given in Fig.3. For both first frequencies (0.4 and 0.8 Hz), there are one or several (respectively) amplification areas along the free surface and in-depth motion decreases regularly down to the bedrock. For the second value (0.8 Hz), the deep deposit also appears more sensitive to some basin edge effects than the shallow basin. For the third frequency (1.2 Hz), we can suspect some little focusing effect since there is a very slow decrease of in-depth motion on the right part of the basin. At the bottom of the deepest part of the basin, there is a rather large value of seismic amplification. The focusing effects is much clearer for frequency 1.4 Hz : in the deepest part of the basin there is a strong increase of in-depth motion. It corresponds to an area of strong motion amplification located inside the alluvial deposit. For larger frequencies (1.8 and 2.0 Hz), there are several parts of the basin where in-depth motion increases. For some places, deep amplification can reach similar values to those obtained along the free surface. At 1.8 Hz, three main areas lead to in-depth motion increase and there are six of them at 2.0 Hz. These results (Fig.3) show a strong influence of focusing effects on in-depth motion amplification. In the next section, we will discuss the dependence of seismic motion on depth by comparing in-depth motion curves for this deep site at various frequencies.

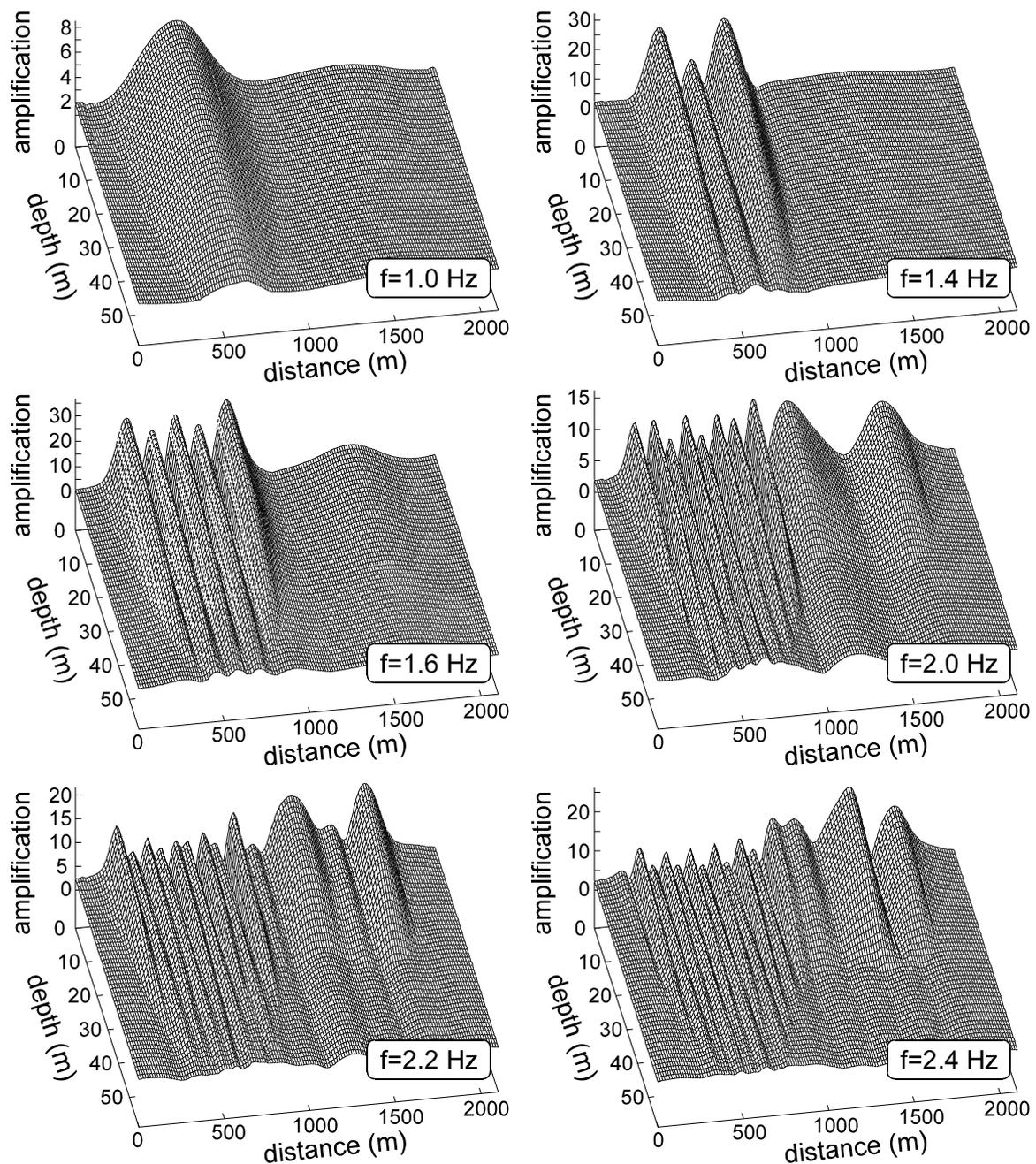

Fig.2 : Amplification in the whole shallow deposit (Nice) at various frequencies.

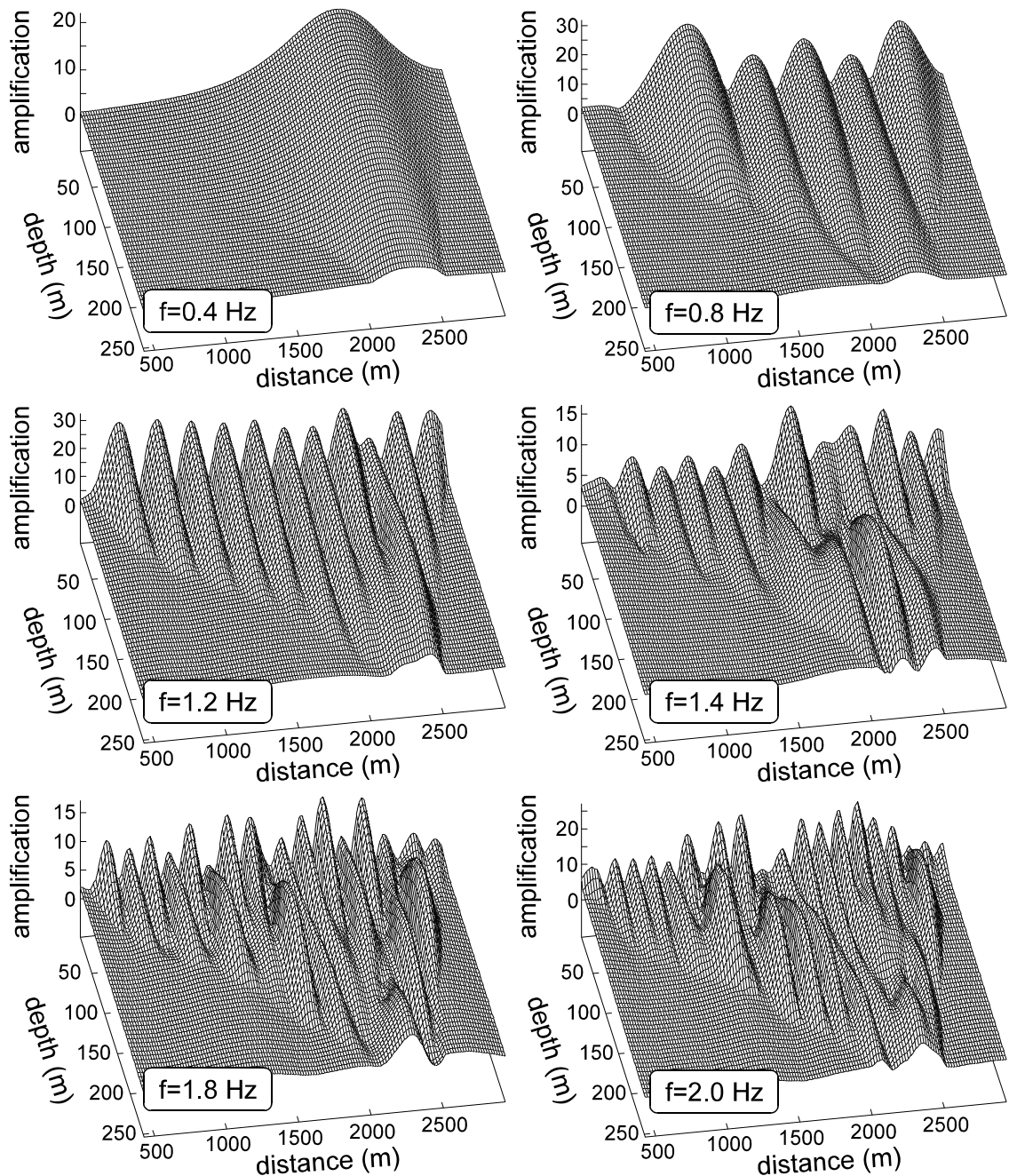

Fig.3 : Amplification in the whole deep deposit (Caracas) at various frequencies.

## 5. Influence on in-depth motion

In the case of the shallow basin, there is always a regular in-depth motion decrease in agreement with the classical rules (Fig.2). For horizontally multilayered media, a simple analytical analysis leads to explicit decreasing laws for in-depth seismic motion [13]. For a two-dimensional shallow regular basin, results given in Fig. 2 follow the same trend than analytical results in the multilayered case.

To investigate the influence of focusing effects on in-depth motion for the deep irregular basin, several curves giving seismic motion versus depth are considered (Fig.4). The variations of in-depth motion are very different. In some places, there could be a strong increase of in-depth motion due to focusing effect. As shown in Fig. 4, at 1.4 Hz, there is a maximum of the seismic motion inside the basin in its deepest part. For larger frequencies (1.8 and 2.0 Hz), the wavelength is shorter and in-depth motion local maxima appear in other parts of the deposit with different energy focusing processes. For the largest frequency (Fig.4), there are even several different large motion areas along the maximum depth. In Fig.4, the local maximum is shown to appear between 200 and 250m. For large frequency values, there are then several related focusing effects corresponding to the focus of seismic waves in shallower areas of the basin at shorter wavelengthes. The focusing effects are then influenced by the geometry of the basin as well as the depth/wavelength aspect ratio.

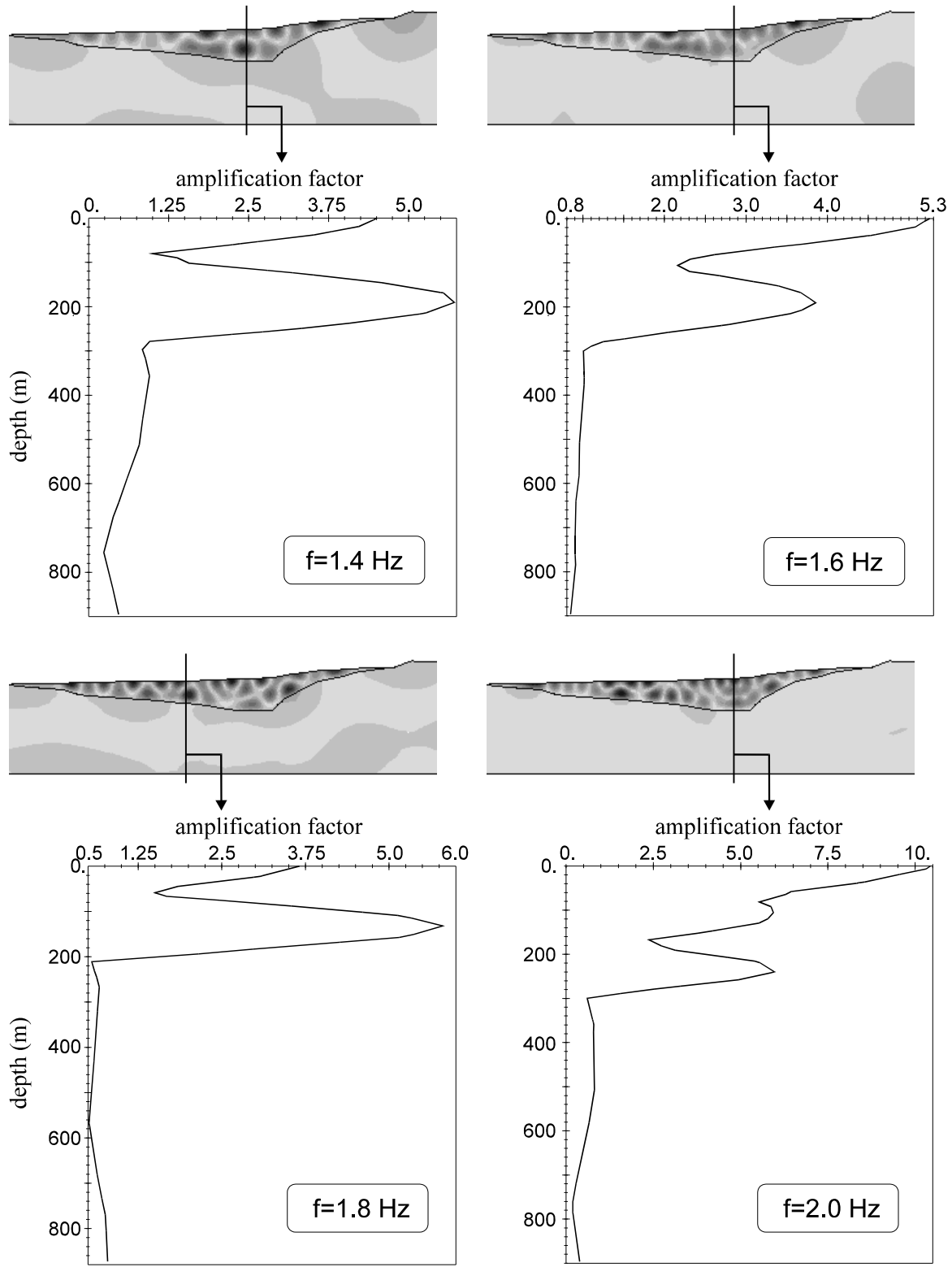

Fig.4 : In-depth motion at various locations and frequencies for the deep deposit (Caracas).

## 6. Conclusion

The analysis of seismic site effects for two very different alluvial basins (shallow, deep) gives interesting results on potential energy focusing effects [8]. For the shallow regular basin considered in Nice, there is no focusing effect and larger amplification is obtained along the free surface. The influence of the basin geometry, vs wavelength, is only observed on the location of maximum amplification areas (deepest part for low frequencies and thinnest part for higher frequencies). For the deep irregular basin in Caracas, various amplification areas are observed inside the basin itself starting in the deepest part of the basin at some intermediate frequency. For larger frequencies (shorter wavelengths), different parts of the basin lead to large deep amplification. In-depth motion variations are consequently influenced by focusing effects. The shallow basin gives a classical decrease of in-depth motion whereas the deep basin can lead to some in-depth motion increases due to energy focusing effects. It is of great interest for the design of earthquake resistant underground structures [12] as well as the analysis of seismic hazard in urban areas [14].